# Survivable MPLS Over Optical Transport Networks: Cost and Resource Usage Analysis

Wojtek Bigos, Bernard Cousin, Stéphane Gosselin, Morgane Le Foll, and Hisao Nakajima

*Abstract*—In this paper we study different options for the survivability implementation in MPLS over Optical Transport Networks (OTN) in terms of network resource usage and configuration cost. We investigate two approaches to the survivability deployment: single layer and multilayer survivability and present various methods for spare capacity allocation (SCA) to reroute disrupted traffic.

The comparative analysis shows the influence of the offered traffic granularity and the physical network structure on the survivability cost: for high bandwidth LSPs, close to the optical channel capacity, the multilayer survivability outperforms the single layer one, whereas for low bandwidth LSPs the single layer survivability is more cost-efficient. On the other hand, sparse networks of low connectivity parameter use more wavelengths for optical path routing and increase the configuration cost, as compared with dense networks. We demonstrate that by mapping efficiently the spare capacity of the MPLS layer onto the resources of the optical layer one can achieve up to 22% savings in the total configuration cost and up to 37% in the optical layer cost. Further savings (up to 9 %) in the wavelength use can be obtained with the *integrated* approach to network configuration over the *sequential* one, however, at the increase in the optimization problem complexity. These results are based on a cost model with different cost variations, and were obtained for networks targeted to a nationwide coverage.

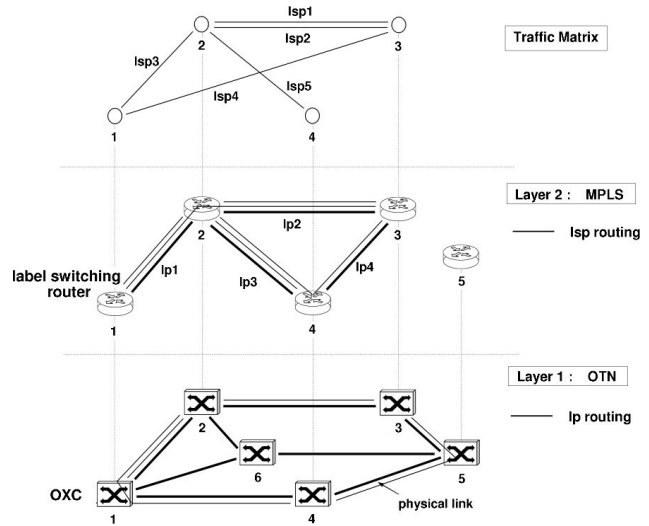

Fig. 1. Architecture of the MPLS over OTN.

## I. INTRODUCTION

THE CONSIDERED MPLS over Optical Transport Network (OTN) represents a multilayer network architecture, where label switching routers (LSR) making up the *MPLS layer* are directly attached to optical cross-connects (OXC) belonging to the *optical layer*. In the optical layer, optical cross-connects are interconnected with point-to-point WDM links in a mesh topology. The interconnection between routers in this architecture is provided by circuit-switched, end-to-end optical channels or *lightpaths* [1]. A lightpath (LP) represents a sequence of fiber links forming a path from a source to a destination router together with a single wavelength on each of these links. The OXCs can switch wavelengths between fiber links without undergoing optoelectronic conversion. A lightpath must be assigned the same wavelength on each link on its route, unless the OXCs support the wavelength conversion capability. The set of lightpaths makes up a *logical network topology*. IP traffic in the form of label switched paths (LSP) is carried in the network over this logical topology using single or multiple *logical hops*. Due to operational simplicity, high capacity potentials offered by WDM transport systems and relatively low implementation cost, as compared with other network technologies (ATM, SDH), the MPLS over OTN is considered as one of the most promising solution for deployment in high capacity backbone networks. Fig. 1 shows an example of the MPLS over OTN architecture.

As network survivability plays a critical role in the network design, a number of recovery schemes have been proposed in the scope of the MPLS over OTN. They are based on two general concepts: *single layer* survivability [2], [3], where recovery mechanisms are implemented only in the MPLS layer and *multilayer* survivability [2], [4], [5], where recovery is employed both in the MPLS and the optical layer. The multilayer survivability has the advantage over the single layer approach in faster and simpler recovery from physical link failures, but it is considered to consume more optical layer resources [2], [4]. This is because with multilayer recovery each network layer reserves some spare resources for rerouting of affected paths, so multiple spare capacity pools are provided, each dedicated to a particular layer. On the other hand, single layer recovery requires more resources from the MPLS layer what may negatively affect the total network configuration cost as these resources are more expensive than the resources of the optical layer.

In this paper we present the design of a survivable MPLS over Optical Transport Network as an integer linear programming (ILP) optimization problem. Our objective is to minimize the amount of network resources used with a given network configuration. There are implemented different methods for spare capacity allocation (SCA) with the single- and multilayer survivability to reroute disrupted traffic. The planning process for SCA is based on two approaches to network configuration: the *sequential* one, where the MPLS layer and the optical layer are planned separately and the *integrated* one, where the whole network is designed in one step. The aspects of spare capacity planning and sequential/integrated approaches to network configuration are related to the design of *multilayer* network architectures and contribute to efficient network configuration in terms of resources usage. The objective of this work is to consider both these aspects in the context of network optimization and to investigate their impact on network resource savings. A set of MPLS over OTN configurations is implemented, where a particular SCA method is combined with a particular configuration approach demonstrating their relative importance to the overall network design in terms of network resource consumption and configuration cost. Despite focusing on specific multilayer network architecture, the collected results show clearly some general tendencies, which make them useful within the overall context of any multilayer network design.

The rest of this paper is organized as follows: Section II presents different SCA methods for the single- and the multilayer survivability implementation. Section III describes two approaches to the MPLS over OTN design: sequential and integrated and explains their impact on the network resource usage. We precise the objectives to be realized with respect to these problems and explain how our work extends the previous studies. In Section IV we define a framework for the survivable MPLS over OTN design. We present algorithms for different spare capacity planning options, define a cost model to be included into the optimization procedure and give exact ILP formulations for the considered problems. Section V concludes with the analysis of the obtained results.

## II. SPARE CAPACITY PLANNING IN MPLS OVER OTN

One of the aspects related to the survivability design is how to allocate spare capacity in a network, so that the total amount of network resources is minimized. The total amount of network resources used with a given SCA option depends on supported failure scenarios and a recovery technique used. Although many recovery schemes can be employed in individual layers of the MPLS over OTN model (e.g. protection vs. restoration, dedicated vs. shared, end-to-end vs. local), we focus here on the multilayer aspects of the survivability design, leaving the problem of the survivability deployment in individual network layers out of consideration. Therefore, only one (and the same) recovery technique is assumed in individual network layers, which is *end-to-end path protection* both with *dedicated* and *shared* spare capacity. It is believed that such a recovery scheme will be always required in the network to protect the integrity of high class services. The considered failure scenarios include physical link failures (e.g. fiber cuts, optical line system failures), *transit*[1] node failures (both router and OXC) and IP/optical interface failures.

With the single layer survivability (Fig. 2(a)) protection is implemented at the LSP level and the MPLS layer covers all failure types. Each *working* LSP (wLSP) has a corresponding *protection* LSP (pLSP), link- and node-disjoint in both network layers. With the multilayer survivability (Fig. 2(b)) protection is implemented both at the LSP level (pLSP) and the lightpath level (pLP). The optical layer protects against physical link and OXC failures, whereas the MPLS layer protects against router and IP/optical interface failures which cannot be detected by the optical layer. The MPLS layer also protects against OXC failures with respect to LSPs transit in the co-located routers. This implies that with the multilayer survivability only *multi-hop* LSPs which are susceptible to router failures have corresponding protection LSPs routed in the MPLS layer (to provide node disjointness against router failures). *Single-hop* LSPs do not require any extra spare capacity from the MPLS layer as they are subject only to the failures resolved at the lightpath level (e.g. $wLSP2$ carried on working lightpath $wLP4$ is protected by protection lightpath $pLP4$). To cover the IP/optical interface failures (e.g. optical line cards, intra-office links and tributary OXC ports), the reach of protection lightpaths is extended towards optical line cards in routers.

Another point to consider with the multilayer survivability implementation is how the MPLS spare capacity used to protect working LSPs is supported by the optical layer. Three options for the spare capacity planning can be considered in this regard [2], [4], and [5]:

1) With a simple capacity planning without any precautions taken, called *double* or *redundant protection* [2], spare capacity in the MPLS layer is protected again in the optical layer. The working LSPs are thus twice protected: once in the MPLS layer and once in the optical layer. This results in an inefficient use of network resources with a small increase in service reliability.

2) An improvement in the optical spare capacity utilization can be achieved by supporting working and protection LSPs on different lightpaths and treating them differently in the optical layer: lightpaths carrying working LSPs are protected while lightpaths supporting protection LSPs are left unprotected (e.g. $wLP3$ in Fig. 2(b) carrying protection LSP $pLSP1$). This option, called LSP '*spare*' *unprotected* [4] requires less resources than double protection; it is however still inefficient in a way that the optical layer still dedicates some resources to support the MPLS spare capacity.

3) Further improvement in spare capacity planning consists in sharing spare resources between the MPLS and the optical layer. With this option, called *interlayer backup resources sharing* (interlayer BRS) [5] or *common pool survivability* [2], [4] the MPLS spare capacity is considered as extra traffic in the optical layer (i.e., carried on unprotected, pre-emptible lightpaths, such as $wLP3$ in

---

[1]Paths originating/terminating at a failed node are considered as lost since they cannot be restored with path protection mechanisms.

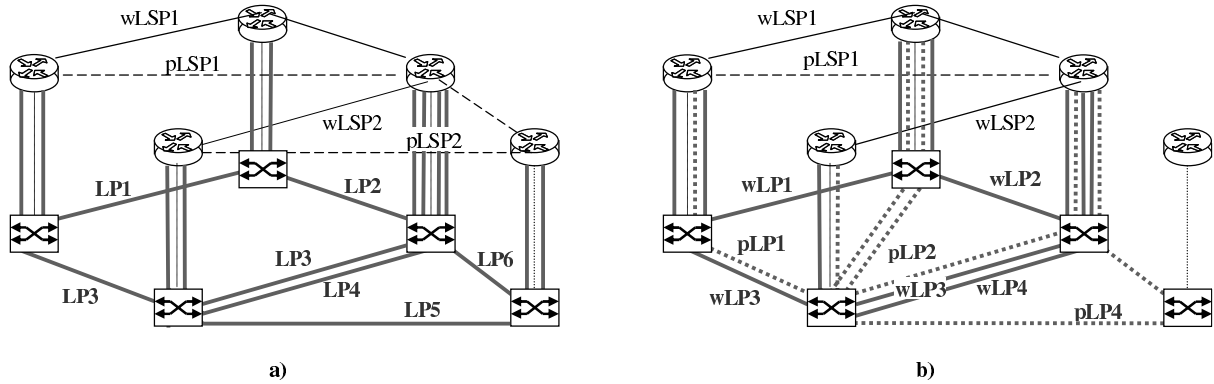

Fig. 2. Two options for the survivability implementation in MPLS over OTN: a) single layer survivability, b) multilayer survivability

Fig. 2(b)). As a consequence, there exists only one spare capacity pool (in the optical layer, for lightpath protection) which can be reused by MPLS recovery schemes when needed (e.g. in Fig. 2(b) protection lightpaths $pLP1$ and $pLP2$ share the wavelengths with working lightpath $wLP3$ carrying protection LSP $pLSP1$). Little or no additional resources in the optical layer are required to support the MPLS spare capacity.

The analysis of the survivability implementation in the MPLS over OTN in terms of resources usage and configuration cost has been already addressed in [4], [6]-[9]. The analysis provided in [6] shows that the single layer recovery is by 10% more cost-effective than the multilayer recovery when lightpaths are not fully utilized with the working traffic, while for the high lightpath utilization it is the opposite. The authors however consider only physical link failures as a possible failure scenario (assuming dual-router architecture to protect against router failures) and the results taking into account more failure scenarios may be different. Various SCA options for the multilayer survivability have been investigated in [4] showing 15% and 20% cost improvements achieved respectively with the "LSP spare unprotected" and the "interlayer BRS" methods over the "double protection." The planning process used in [4] does not guarantee the recovery from the OXC failures when using the interlayer BRS option. This work extends the previous studies by adding more failure scenarios, including physical link failures, node failures (both router and OXC) and IP/optical interface failures. Exact planning processes are given for different SCA options and a cost model is defined which allows the network configuration cost to be modified according to the price evolution of network components. Finally, we present the integrated approach to the network design where the MPLS and optical layer are optimized jointly, leading to extra savings in network resources.

### III. SEQUENTIAL VS. INTEGRATED APPROACH TO THE MPLS OVER OTN CONFIGURATION

In the MPLS over OTN both network layers can be combined using either the *overlay* or the *peer* interconnection model [10]. In the overlay model, the MPLS and the optical layer are controlled separately and each layer has its own instance of the control plane. There are two separate routing processes in the network: the MPLS layer routes LSPs in the logical topology using either existing lightpaths or requesting a new lightpath *directly* connecting LSP endpoints from the optical layer; then the optical layer routes the lightpaths in the physical topology. In the peer model, a single control plane controls both the MPLS and the optical layer. There is one routing process which runs across both layers and logical and physical links are considered jointly in route selection. As a result, the routing process can use some existing ligtpaths and simultaneously create additional lightpaths on physical links to achieve the most optimal route selection.

Most previous studies on routing implementation either with the overlay [3] or the peer interconnection model [11]-[13] concentrate exclusively on analyzing the blocking probability of a given routing approach under dynamic traffic conditions. There are however no analyses showing their performance in terms of network resources consumption. Our objective is to implement the concepts of sequential and integrated routing as network optimization problems and to compare the use of network resources achieved with both methods. The considered network resources include:

- In the MPLS layer:
- Amount of packet processing in routers which is proportional to the volume of the transit traffic at each router (i.e. neither originating nor terminating in a router); by minimizing it we improve the router throughput and minimize the packet queuing delay.
- Number of IP/optical interfaces in routers which constitute a significant part of the configuration cost;
- In the optical layer:
- Number of wavelengths and wavelength-switching equipment used to route a given set of lightpaths in the physical topology.

With the sequential approach to network configuration, as presented in Fig. 3(a), the optimization problem consists of two sub-problems. The first sub-problem takes as an input the traffic matrix in terms of LSPs to be routed in the network and returns as a result the set of lightpaths to be established in the optical layer (i.e. the logical topology) and the routing of LSPs over the logical topology. Thus, only a part of the MPLS over OTN configuration is solved by the first sub-problem and the optimization function takes as an objective minimizing *only* the IP/MPLS layer resources, i.e. the total number of lightpaths between all node pairs $\sum_{(i,j)} w_{(i,j)}$ and the total transit

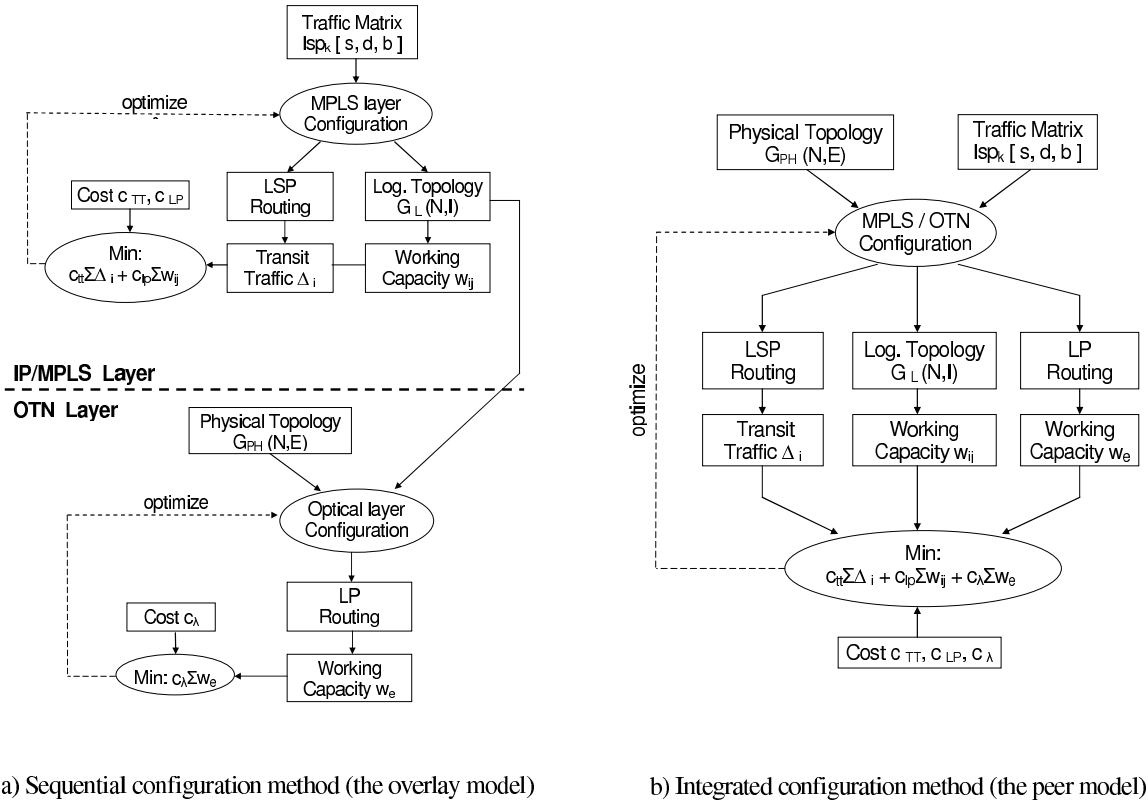

a) Sequential configuration method (the overlay model)    b) Integrated configuration method (the peer model)

Fig. 3. Sequential vs. integrated approach to the MPLS over OTN configuration. Legend: traffic matrix - set of $K$ LSPs; each LSP has associated (s)ource and (d)estination point, and (b)andwidth; $G_{PH}(N, E)$ - graph representing the physical network topology of $N$ vertices (nodes) and $E$ edges (physical links); $G_L(N, I)$ - graph representing the logical topology of $N$ nodes and $I$ logical links; $w_{i,j}$ - capacity per node pair in the MPLS layer, measured in the number of lightpaths installed between the node pair $i - j$; $\Delta_i$ - transit traffic processed by node $i$; $w_e$ - capacity per link in the optical layer, measured in the number of wavelengths used on the physical link $e$ to carry lightpaths; $c_{LP}$ - lightpath cost; $c_\lambda$ - wavelength cost; $c_{TT}$ - transit traffic cost.

traffic $\sum_i \Delta_i$ processed by routers. The second sub-problem takes the set of lightpaths to be established on physical links and the physical network topology as input parameters and returns the routing of lightpaths in the physical topology optimizing *only* resources of the optical layer, i.e. the total number of wavelengths $\sum_e w_e$ used to route the lightpaths in the physical topology. The lightpath cost $c_{LP}$, the wavelength link cost $c_\lambda$ and the transit traffic cost $c_{TT}$ are included into the optimization procedure to account for the total configuration cost of the network. Note, however, that such decomposition is approximate or inexact. Solving the sub-problems in sequence and combining the solutions may not result in the optimal solution for the fully integrated problem. On the other hand, with the integrated approach, as depicted in Fig. 3(b), there is only one optimization problem implemented which provides a full MPLS over OTN design in one step. The optimization function used to control the configuration process explicitly accounts for the resource usage both in the MPLS and the optical layer. The global optimization of network resources is thus possible with the integrated approach. Nevertheless, achieving an absolutely optimal solution of such a problem may be hard as the computational complexity of the algorithm solving the network configuration problem in a combined fashion is high.

## IV. PROBLEM FORMULATION

We consider the following network design problem. Given the offered traffic matrix (in terms of static LSP connections), the physical network topology and a set of constraints on logical and physical link capacities, we search for an MPLS over OTN configuration which (i) provides 100% restorability against the considered failure scenarios and (ii) minimizes the total resource usage in both network layers. A cost model is included into the optimization procedure which represents the monetary cost of various network components. Thereby, by minimizing the amount of network resources the cost of a given network configuration is also optimized. We use integer linear programming (ILP) as an optimization technique and formulate the problem using linear models. The problem solution provides a complete specification to the logical topology design and routing of working and protection paths both in the MPLS and the optical layer together with the resource usage at the minimal cost.

We make the following assumptions in our study:
1) The traffic matrix is symmetric and the lightpaths are bidirectional. Two lightpaths from a pair are routed over the same physical route but in opposite directions.
2) The optical layer has an *opaque* configuration with photonic OXCs (i.e. which switch wavelengths optically) surrounded by WDM transponders performing OEO conversion. Transponders perform signal regeneration and adaptation functions including wavelength conversion.

3) As the optical layer supports wavelength conversion, the wavelength continuity constraint is not considered (under which a lightpath is assigned the same wavelength on all links on its route). This assumption reduces the problem in terms of ILP variables and constraints and makes it more computationally tractable.

### A. Survivability Implementation

Algorithms for the MPLS over OTN design using the sequential configuration approach combined with various options for the survivability implementation are presented in Fig. 4(a)–(d). Each algorithm consists of four planning processes (steps): two for the network design with *working* paths, respectively in the MPLS and the optical layer (steps I and III), and two others for the network configuration with *protection* paths (steps II and IV). Each part is implemented as a separate optimization problem using a distinct ILP formulation. The planning process for the computation of working LSPs and lightpaths (i.e. steps I and III in each algorithm) is analogous to the one presented in Fig. 3(a). Next, the routing of protection paths is determined for the set of pre-computed working paths and spare capacity is allocated. With the single layer survivability (see Fig. 4(a)), the spare capacity planning is done only in the MPLS layer (in step II). The planning process takes as the inputs the considered failure scenarios and the routing of working LSPs computed in step I. Then, the routing of protection LSPs is computed taking into account the constraints for protection routing in the MPLS layer (see below). The optimization function aims at minimizing the total transit traffic $\sum_i \Delta_i$ and the spare capacity $\sum_{(i,j)} s_{(i,j)}$ consisting of the lightpaths carrying protection LSPs. With the multilayer survivability implementation (see Fig. 4(b)–(d)), the spare capacity planning is done both in the MPLS and the optical layer with the objective to minimize the resources of each layer. Contrary to the single layer survivability, only multi-hop LSPs are subject to the protection LSP routing in step II; single hop LSPs are protected at the lightpath level. With the "double protection" method (Fig. 4(b)), both the lightpaths carrying working and protection LSPs are protected in the optical layer. With the "LSP spare unprotected" option (Fig. 4(c)), working and protection LSPs are routed over two disjoint sets of lightpaths (sets $G_{L1}$ and $GL_2$) and only the lightpaths carrying working LSPs (from $G_{L1}$) have protection lightpaths; the lightpaths carrying protection LSPs (from $G_{L2}$) are not protected. With the "interlayer BRS" (Fig. 4(d)), not only the lightpaths carrying protection LSPs are unprotected but they also share the wavelengths with the protection lightpaths to optimize further the wavelength use (i.e. spare capacity $s_e$ is shared with the working capacity $w_{e2}$ supporting the lightpaths which carry protection LSPs).

Next, the planning process was repeated but using the integrated configuration approach where the routing of working LSPs and the corresponding lightpaths (i.e. steps I, III) was implemented as a single optimization problem according to the scheme presented in Fig 3(b). The same methodology was then used for protection LSPs and the lightpaths carrying protection LSPs (steps II and III). Note, however, that to combine the routing of working and protection LSPs and lightpaths (i.e. steps I–II and III–IV) we use the sequential approach only, i.e. routing of protection paths is determined for the set of pre-computed working paths. The results from computing working and protection paths jointly are reported in [20] showing 8%-12% savings in spare capacity than if the paths are computed separately, however, at the increase in the optimization problem complexity.

The following rules are defined for the protection routing, i.e., how to route protection paths, so that all the considered failure scenarios are supported:

- For the single layer survivability:
  - Each working LSP ($wLSP$) has a protection LSP ($pLSP$); corresponding working and protection LSPs are node- and link disjoint both in the logical and physical topology.
- For the multilayer survivability:
  - Each *multihop* working LSP has a protection LSP; corresponding working and protection LSPs are link- and node-disjoint in the logical topology.
  - With the "double protection" method each lightpath has a protection lightpath, with the "LSP spare unprotected" and "interlayer BRS" method only the lightpaths carrying working LSPs have corresponding protection lightpaths. Respective working and protection lightpaths are link- and node-disjoint in the physical topology.

Additional requirement for the "LSP spare unprotected" and "interlayer BRS" methods:

- Corresponding working and protection LSPs are node disjoint in the physical topology; this prevents a working LSP and its protection LSP from failing simultaneously in case of an OXC failure.

And for the "interlayer BRS":

- Lightpaths transiting an OXC and LSPs transiting the co-located router are protected on different physical links; this prevents the failed entities from competing for the same spare resources in case of an OXC failure.

### B. Cost Model

By including the cost of various network elements into the optimization procedure we optimize the network configuration cost which depends on the number and type of established communication channels. In this study we do not deal with the investments for the initial network deployment which include the cost of lying/leasing fibers and the cost of WDM line systems (without transponders), i.e. WDM (de)multiplexers and optical amplifiers. As it has been stated in the problem formulation, the initial network topology determining these costs is given to the problem as an input parameter.

The following cost components are included into the optimization procedure: the cost of IP/optical interface cards in routers $c_{P\_IP}$, the cost of OXC ports $c_{P\_OXC}$ and the cost of optical transponders $c_{TR}$. The costs of the router and OXC equipment are incorporated, respectively, into the IP/optical interface cost $c_{P\_IP}$ and the OXC port cost $c_{P\_OXC}$. Additional cost $c_{TT}$ is associated with the amount of the transit traffic processed by routers as a penalty for diminishing the packet processing capability of a router which could be otherwise used by the originating/terminating traffic. The cost $c_{TT}$ is specified by the $c_{P\_IP}$ cost per traffic unit:

$$c_{TT} = \frac{c_{P\_IP}}{C} \times \text{ transit traffic} \quad (1)$$

where $C$ denotes the IP/optical interface card rate.

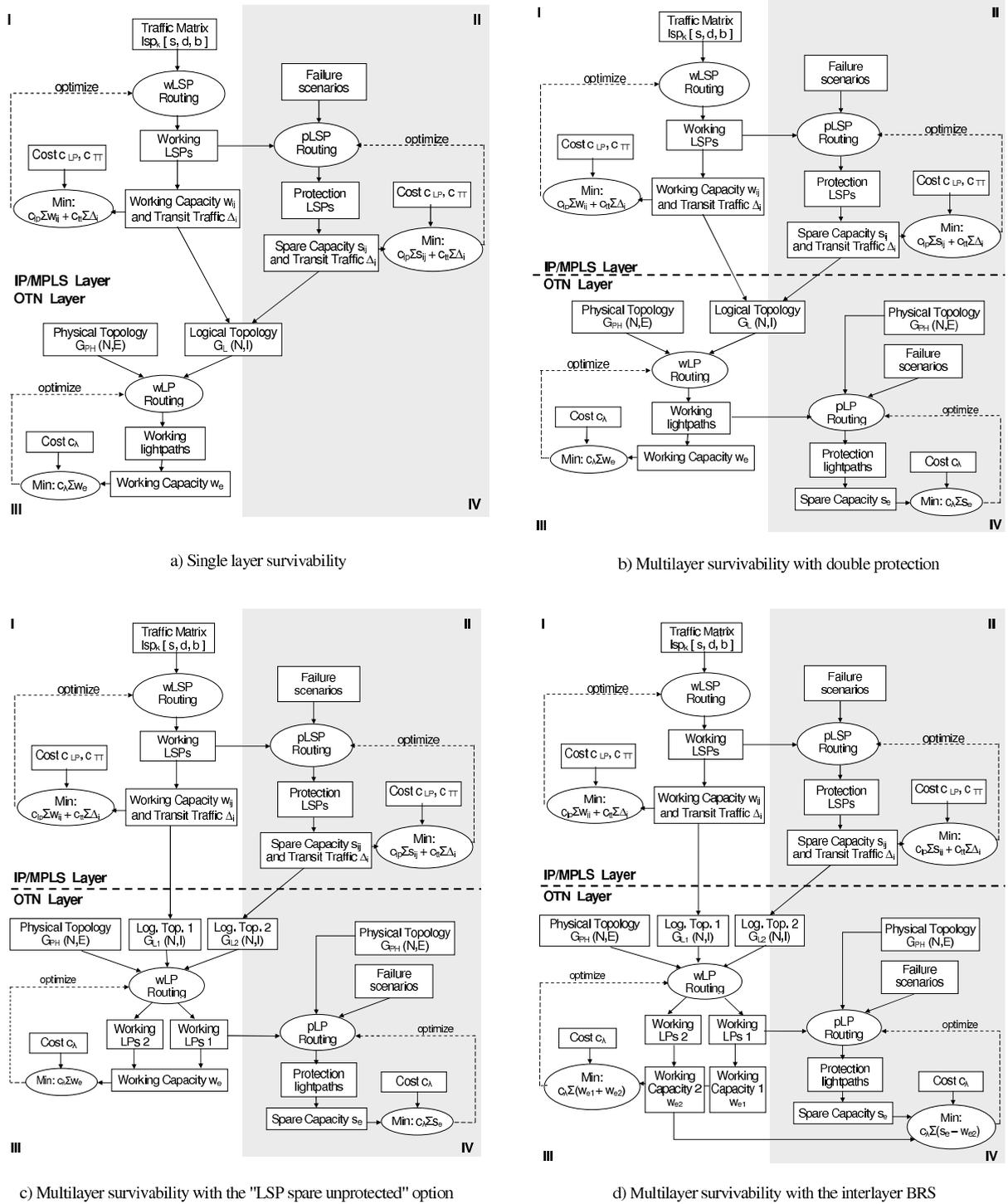

Fig. 4. Algorithms for the MPLS over OTN design using the sequential approach with various options for the survivability implementation. Legend: $s_{i,j}$ - spare capacity per node pair in the MPLS layer, measured in the number of lightpaths carrying protection LSPs (pLSP) between the node pair $i-j$; $s_e$ - spare capacity in the optical layer, measured in the number of wavelengths used on the physical link $e$ to carry protection lightpaths (pLP); The other symbols have the meaning as specified in Fig. 3.

The costs of network elements are mutually related according to three different cost ratios (CR), as presented in Table I. Ratio CR_1, represents the current prices of the elements (year 2006) for 10 Gbps IP/optical interfaces and transponders, $256 \times 256$ port *photonic* OXCs and 200 Gbps routers. While CR_1 represents the current trend of predominance of the optical transmission cost (i.e. IP/optical interface cost $c_{P\_IP}$) within the overall cost structure, ratios CR_3 and CR_2 account for, respectively, optical switching and regeneration/adaptation costs ($c_{P\_OXC}$, $c_T$) as predominant factors.

TABLE I
COST RATIOS OF NETWORK ELEMENTS FOR DIFFERENT COST SCENARIOS

| Network Element | Cost Scenario | | |
|---|---|---|---|
| | CR_1 | CR_2 | CR_3 |
| WDM transponder $c_T$ | 1 | 8 | 0,5 |
| IP/optical interface card $c_{P\_IP}$ | 8 | 0,5 | 1 |
| Photonic OXC port $c_{P\_OXC}$ | 0,5 | 1 | 8 |

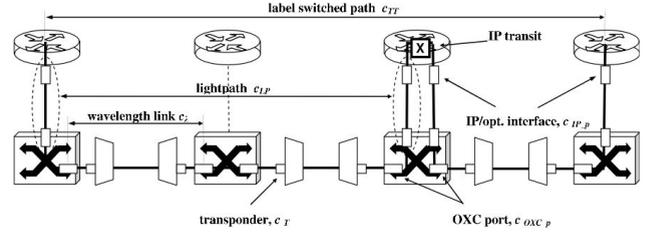

Fig. 5. An example of the MPLS over OTN configuration combined with the cost of network elements.

The total cost of a network configuration is a sum of the transit traffic cost and the cost of individual circuits: lightpaths and wavelength links, as depicted in Fig. 5. It is assumed that each wavelength link requires 2 transponders and 2 OXC ports (i.e. 1 transponder and 1 OXC port for each termination point) and each lightpath (protection and working) requires 2 IP/optical interface cards and 2 OXC ports. Thus, the cost components $c_{LP}$ and $c_\lambda$ presented in schemes in Fig. 4(a)–(d) are as follows: $c_{LP} = 2(c_{P\_IP} + c_{P\_OXC}), c_\lambda = 2(c_{P\_OXC} + c_{TR})$.

### C. ILP Formulation

The ILPs are defined using the *node-arc* formulation, where network nodes are indexed by subscripts and edges are specified by an $(x,y)$ node-name pair:
- $s$ and $d$ denote source and destination nodes of an LSP,
- $i$ and $j$ denote originating and terminating nodes in a lightpath,
- $q$ denotes a $q-th$ lightpath between the $(i,j)$ pair,
- $m$ and $n$ denote endpoints of a physical link.

*Inputs:*

N     set of nodes; each element represents a generic network node being a combination of a router and the co-located OXC.

LSP     set of $K$ LSPs to be routed in the network; each element $lsp_k$ represents an indivisible traffic flow to be routed on a single LSP and has associated a triple: $\{s(lsp_k), d(lsp_k), b(lsp_k)\}$ denoting respectively its source, destination node and bandwidth.

LP     set of lightpaths determining the logical topology; each element $lp$ has associated a triple: $\{i(lp), j(lp), q(lp)\}$ denoting respectively its originating, terminating node and the multiple.

p_LSP     set of LSPs to be protected; a subset of LSP.

p_LP     set of lightpaths to be protected; a subset of LP.

N_ex [$lsp$]     for each $lsp$ in p_LSP, set of nodes to be excluded from the route of its protection LSP to respect the protection routing constraints.

N_ex [$lp$] for each lightpath in LP, set of nodes to be excluded from its route to respect the protection routing constraints. Note, that in general not only protection but also working lightpaths are subject to protection routing, in particular, to provide disjointness between (corresponding) working and protection LSPs in the physical topology with the single layer survivability.

$P_{N \times N}$     physical topology matrix, where the element $p_{(m,n)} = p_{(n,m)} = 1$ if there exist a physical link between nodes $m$ and $n$; otherwise $p_{(m,n)} = p_{(n,m)} = 0$, (i.e. physical links are bidirectional). It is assumed that there are no multiple links between node pairs.

C     capacity of a lightpath.

Q     max. number of lightpaths of a given status (i.e. *working*, *protection*) to be setup between a given node pair. It is assumed that $q \in \{1, 2\}$.

T     max. number of IP/optical interfaces in a router.

W     max. number of wavelengths on a physical link.

$c_{TT}$     cost of the transit traffic per traffic unit.

$c_{LP}$     cost of setting up a lightpath.

$c_\lambda$     cost of allocating a wavelength.

Variables:

1) Logical topology variables $w\beta_{(i,j),q}, p\beta_{(i,j),q} = 1$ if there exist a $q-th$ lightpath from node $i$ to node $j$ carrying, respectively, working and protection LSPs; 0 otherwise.
2) LSP routing variables $w\delta_{(i,j),q}^{lsp}, p\delta_{(i,j),q}^{lsp}$ if, respectively, the working / protection LSP $lsp$ is routed on the $q-th$ lightpath from node $i$ to node $j$; 0 otherwise.
3) Lightpath routing variables:
a. With the sequential configuration method: $w\lambda_{(m,n)}^{lp}, p\lambda_{(m,n)}^{lp}$ if, respectively, the working / protection lightpath $lp$ is routed on the physical link $(m, n)$; 0 otherwise.
b. With the integrated configuration method: $w\lambda_{(m,n)}^{(i,j),q}$, if a $q-th$ working lightpath from node $i$ to node $j$ is routed on the physical link $(m, n)$; 0 otherwise. Note, that as protection lightpaths are routed using the sequential method only, no protection routing variable is defined here.

*Optimization function:*

The objective is to limit the total network resource usage by minimizing costs of the transit traffic and capacity allocation (working + spare) in both network layers:

$$\text{Minimize: } c_{TT} \cdot \sum_n \Delta_n + c_{LP} \cdot \sum_{(i,j)} \left(w_{(i,j)} + s_{(i,j)}\right) + c_\lambda \cdot \sum_{(m,n)} \left(w_{(m,n)} + s_{(m,n)}\right) \quad (2)$$

where:

$$\Delta_n = \sum_{lsp} b\_req(lsp) \cdot \sum_i \sum_q \left( w\delta^{lsp}_{(i,n),q} + p\delta^{lsp}_{(i,n),q} \right) - \sum_{lsp:d(lsp)=n} b\_req(lsp) \qquad (3)$$

$$w_{(i,j)} + s_{(i,j)} = \sum_q w\beta_{(i,j),q} + \sum_q p\beta_{(i,j),q} \qquad (4)$$

With the sequential configuration approach:

$$w_{(m,n)} + s_{(m,n)} = \sum_{lp} w\lambda^{lp}_{(m,n)} + \sum_{lp} p\lambda^{lp}_{(m,n)} \qquad (5)$$

With the integrated configuration approach:

$$w_{(m,n)} + s_{(m,n)} = \sum_{(i,j)} \sum_q w\lambda^{(i,j),q}_{(m,n)} + \sum_{lp} p\lambda^{lp}_{(m,n)} \qquad (6)$$

The cost components $c_{TT}$, $c_{LP}$ and $c_\lambda$ are associated as specified in the cost model (see Sec. IV-B)

*Constraints*:

1) Constraints for the logical topology design:

$$\sum_{j:j\neq i} \sum_q w\beta_{(i,j),q} + p\beta_{(i,j),q} \leq T, \qquad \forall i \in N \qquad (7)$$

$$\sum_{i:i\neq j} \sum_q w\beta_{(i,j),q} + p\beta_{(i,j),q} \leq T, \qquad \forall j \in N \qquad (8)$$

2) Constraints for the LSP routing:

$$\sum_{j:j\neq i} \sum_q w\delta^{lsp}_{(i,j),q} - \sum_{j:j\neq i} \sum_q w\delta^{lsp}_{(j,i),q}$$
$$= \begin{cases} 1, & \text{if } i = s(lsp) \\ -1, & \text{if } i = d(lsp) \\ 0, & \text{otherwise} \end{cases} \qquad (9)$$
$$\forall lsp \in LSP, \quad i \in N$$

$$\sum_{\substack{j:j\neq i \\ j\notin N\_ex[lsp]}} \sum_q p\delta^{lsp}_{(i,j),q} - \sum_j \sum_q p\delta^{lsp}_{(j,i),q}$$
$$= \begin{cases} 1, & \text{if } i = s(lsp) \\ -1, & \text{if } i = d(lsp) \\ 0, & \text{otherwise} \end{cases} \qquad (10)$$
$$\forall lsp \in LSP, \quad i \notin N\_ex[lsp]$$

$$w\delta^{lsp}_{(i,j),q} + p\delta^{lsp}_{(i,j),q} \leq 1$$
$$\forall lsp \in p\_LSP, \quad (i,j) \in N^2, \quad q \qquad (11)$$

$$\sum_{lsp} b\_req(lsp) \cdot w\delta^{lsp}_{(i,j),q} \leq w\beta_{(i,j),q} \cdot C$$
$$\forall (i,j) \in N^2, \quad q \qquad (12)$$

$$\sum_{lsp} b\_req(lsp) \cdot p\delta^{lsp}_{(i,j),q} \leq p\beta_{(i,j),q} \cdot C$$
$$\forall (i,j) \in N^2, \quad q \qquad (13)$$

3) Constraints for the lightpath routing:

With the sequential configuration approach:

$$\sum_{\substack{m:m\neq n \\ m\notin N\_ex[lsp]}} w\lambda^{lp}_{(n,m)} - \sum_m w\lambda^{lp}_{(m,n)}$$
$$= \begin{cases} 1, & \text{if } n = i(lsp) \\ -1, & \text{if } n = j(lsp) \\ 0, & \text{otherwise} \end{cases} \qquad (14)$$
$$\forall lp \in LP; \quad i \notin N\_ex[lp]$$

$$\sum_{\substack{m:m\neq n \\ m\notin N\_ex[lsp]}} p\lambda^{lp}_{(n,m)} - \sum_m p\lambda^{lp}_{(m,n)}$$
$$= \begin{cases} 1, & \text{if } n = i(lsp) \\ -1, & \text{if } n = j(lsp) \\ 0, & \text{otherwise} \end{cases} \qquad (15)$$
$$\forall lp \in p\_LP; \quad n \notin N\_ex[lp]$$

$$w\lambda^{lp}_{(m,n)} + p\lambda^{lp}_{(m,n)} \leq 1 \quad \forall lp \in p\_LP; \quad (m,n) \notin N^2 \qquad (16)$$

$$\sum_{lp} w\lambda^{lp}_{(m,n)} + p\lambda^{lp}_{(m,n)} \leq W \quad \forall (m,n) \notin N^2 \qquad (17)$$

With the integrated configuration approach:

$$\sum_{\substack{m:m\neq n \\ m\notin N\_ex[lsp]}} w\lambda^{(i,j),q}_{(n,m)} - \sum_m w\lambda^{(i,j),q}_{(m,n)}$$
$$= \begin{cases} w\beta_{(i,j),q}, & \text{if } n = i \\ -w\beta_{(i,j),q}, & \text{if } n = j \\ 0, & \text{otherwise} \end{cases} \qquad (18)$$
$$\forall (i,j) \in N^2, \quad n \notin N\_ex[lp], \quad q$$

$$w\lambda^{(i,j),q}_{(m,n)} + p\lambda^{lp}_{(m,n)} \leq 1 \quad \forall (i,j), (m,n) \in N^2, q,$$
$$p \in p\_LP : i(lp) = i,$$
$$j(lp) = j, q(lp) = q \qquad (19)$$

$$\sum_{(i,j)} \sum_q w\lambda^{(i,j),q}_{(m,n)} + \sum_{lp} p\lambda^{lp}_{(m,n)} \leq W \quad \forall (m,n) \notin N^2 \qquad (20)$$

4) Binary constraints:

$$w\beta_{(i,j),q}, p\beta_{(i,j),q},$$
$$w\delta^{lsp}_{(i,j),q}, p\delta^{lsp}_{(i,j),q},$$
$$w\lambda^{lp}_{(m,n)}, p\lambda^{lp}_{(m,n)}, w\lambda^{(i,j),q}_{(m,n)} \in \{0,1\} \qquad (21)$$

Eq. (3) specifies the transit traffic at node $i$ as a difference between the total incoming traffic and the traffic terminated at $i$. Eq. (4) specifies the working and spare capacity per node pair in the MPLS layer as a sum of lightpaths carrying working and protection LSPs between a general node pair $(i,j)$. Eqs. (5) and (6) determine the working and spare capacity per link in the optical layer as a sum of wavelengths carrying working and protection lightpaths on a general link $(m,n)$. Eqs. (7) and (8) make the number of lightpaths originating and terminating at node $i$ does not exceed $T$. For each LSP (working +protection), Eqs. (9) and (10) specify the flow conservation constraint at every node on its route. Some nodes are excluded from the routing of protection LSPs (Eq. (10)) to meet the protection routing constraints. Eq. (11) provides logical link disjointness between corresponding working and protection LSPs. Eqs. (12) and (13) make the total traffic carried by all LSPs on the lightpath $(i,j)$, $q$ not to exceed the lightpath capacity. For each OXC and lightpath (working + protection), Eqs. (14), (18), and (15) specify the flow conservation constraint at the lightpath level. It states that the number of lightpaths incoming to and outgoing from a node is equal. With the integrated approach the logical topology and lightpath routing are determined jointly (cf. Eq. (18)), whereas with the sequential approach lightpath routing is determined for the set of pre-computed lightpaths (cf. Eq. (14)). Eqs. (16) and (19) provide link-disjointness between corresponding working and protection lightpaths. Eqs. (17) and (20) make the number of lightpaths routed on the physical link $(m,n)$ not to exceed the total number of wavelengths. Binary constraint (21) ensures that the variables take only 0/1 values.

TABLE II
RUNNING TIMES OF THE ILP ALGORITHMS FOR DIFFERENT PROBLEM SIZES (VALUES IN BRACKETS DENOTE ACHIEVED OPTIMALITY GAPS)

| Computation of working paths | | | | | |
|---|---|---|---|---|---|
| Config. Approach | Problem size (N) | | | | |
| | 8 | 10 | 12 | 14 | 15 |
| Sequential | 12.4 s (0.77%) | 70 s (0.72%) | 951 s (0.5%) | 1403 s (0.41%) | 3632 s (0.83%) |
| Integrated | 101 s (1.62%) | 542 s (1.66%) | 1211 s (1.18%) | 18000 s (2.8%) | unsolved |

| Computation of protection paths | | | | | |
|---|---|---|---|---|---|
| Config. Approach | Problem size (N) | | | | |
| | 8 | 10 | 12 | 14 | 15 |
| Sequential | 70 s (optimal) | 1011 s (0.44%) | 2542 s (0.66%) | 17093 s (0.32%) | unsolved |
| Integrated | 550 s (0.34%) | 8856 s (1.12%) | 18000 s (1.76%) | unsolved | ----- |

In a network of $N$ nodes, $E$ links and max. $q$ lightpaths per node pair supporting a traffic matrix composed of $K$ LSPs, the size of the problem solved with the sequential approach is $\approx q \cdot N^2 \cdot K/2$ in terms of variables whereas the same problem solved with the integrated approach is $\approx q \cdot N^2 \cdot (K/2 + E)$.

## V. NUMERICAL RESULTS

The problems specified above were implemented and solved using the CPLEX 9.0 optimization package. All experiments were carried out on a HP Alpha workstation with a 1 GHz CPU and 2 GB RAM running Tru64 UNIX OS. The system parameters were set as follows: $C = 10$ Gbps, $W = 32$, $Q = 2, T = 2Q \cdot (N-1)$. The number of wavelengths per link $W$, IP/optical interface cards per node $T$, the router throughput and the OXC size were over provisioned in a way that no resources shortage constraints were affecting the configuration process, but only the optimization function used.

### A. Computational Efficiency

The logical topology design problem belongs to the class of NP-hard problems for which no efficient (i.e. polynomial time) algorithms are known. It is therefore essential to verify the computational limits of the proposed ILP-based solution method. The problem complexity measure was the execution time of the algorithms as a function of the problem size (in terms of the number of nodes and traffic demands). Solution times of the network configuration based on the sequential and integrated approaches were compared for different problem sizes and for the computation of working vs. protection paths. The solver was set to stop any optimization within a maximum time limit of 5 hours and a 3% solution optimality gap was assumed. If in any case the solver stopped without full termination, the solution quality achieved so far has been reported in terms of the optimality gap. The results are summarized in Table II. The obtained results show that by setting reasonable optimality gaps and run-time limits on ILP algorithms quite good solutions to the specified problems can be obtained even without a full termination, at least for moderate size networks (up to 15 nodes with fully meshed traffic matrices). Nevertheless, the use of heuristics is inevitable to attack larger networks. Some well-known heuristics for computing a set of disjoint paths, such as, for example, branch exchange described in [9] or others, based on Evolutionary Algorithms and presented in [8, 20-21] can be used within the presented design framework for the SL and ML survivability to improve the scalability of the proposed solutions.

The solution times of the optimization based on the integrated approach is about one order of magnitude longer than when using the sequential approach and the difference in solution times for the two methods increases as the problem size grows. This was expected as in a network of $N$ nodes, $E$ links and max $q$ lightpaths per node pair the complexity of the integrated approach increases by $\sim q \cdot E \cdot N^2$ faster in terms of variables and constraints, as compared to the sequential approach. As a consequence, for the biggest solved problems the solution could only be obtained using the sequential approach. Solution times for the computation of protection paths were longer due to additional constraints for protection routing: (11) and (15).

### B. Analysis of Results

For the survivability implementation we used a test network targeted to a nationwide coverage, presented in Fig. 6. The network has a bi-connected mesh topology consisting of 12 nodes and 24 physical links. There are 10 nodes distributing the nationwide traffic (1-10) and 2 gateways providing the Internet access (11, 12). The traffic matrix is population-weighted and consists of 56 bidirectional, symmetric traffic demands or 126 LSPs (as some demands consist of more than 1 LSP). To check for different values of the LSP bandwidth and the total offered traffic, three groups of tests were carried out: for the average LSP bandwidth equal to 2.5, 5.0 and 7.5 Gbps. The results obtained with the sequential configuration method are summarized in Table III(a)–(c) and in Fig. 7.

*1) Single layer vs. multilayer survivability analysis.*

TABLE III
NETWORK RESOURCE USAGE AND THE CONFIGURATION COST FOR DIFFERENT SURVIVABILITY OPTIONS. THE NUMBER OF LIGHTPATHS IN BRACKETS (ROW 2) DENOTES THE LIGHTPATHS CARRYING PROTECTION LSPS. THE NUMBER OF WAVELENGTHS IN BRACKETS (ROW 3) DENOTES EXTRA WAVELENGTHS NEEDED TO ACCOMMODATE PROTECTION LSPS WITH THE "INTERLAYER BRS." THE COST PERCENTAGE DENOTES THE COST SAVINGS WITH RESPECT TO THE MOST EXPENSIVE METHOD.

(a) 126 LSPs; LSP Bandwidth=1, 2, 3 Gbps; Tot. Traffic=300Gbps

|  | Single layer Survivability | Multilayer Survivability | | |
| --- | --- | --- | --- | --- |
|  |  | double protection | LSP spare unprotected | interlayer BRS |
| Transit traffic | 124 Gbps | 94 Gbps | 92 Gbps | 92 Gbps |
| # of lightpaths | 56 (33) | 82 (18) | 66 (21) | 66 (21) |
| # of wavelengths | 140 | 172 | 138 | 108 (4) |
| Total cost | 1471 (26%) | 1985 | 1626 (18%) | 1537 (22%) |
| Optical layer cost | 420 (19%) | 516 | 414 (25%) | 324 (37%) |

(b) 126 LSPs; LSP Bandwidth=2, 4, 6 Gbps; Tot. Traffic=600Gb/s

|  | Single layer Survivability | Multilayer Survivability | | |
| --- | --- | --- | --- | --- |
|  |  | double protection | LSP spare unprotected | interlayer BRS |
| Transit traffic | 262.5 Gbps | 100 Gbps | 97.5 Gbps | 97.5 Gbps |
| # of lightpaths | 143 (79) | 160 (16) | 148 (20) | 148 (20) |
| # of wavelengths | 329 | 334 | 297 | 267 (7) |
| Total cost | 3628 (5%) | 3802 | 3485 (8%) | 3395 (11%) |
| Optical layer cost | 987 (1.5%) | 1002 | 891 (11%) | 801 (20%) |

(c) 126 LSPs; LSP Bandwidth=3, 6, 9 Gbps; Tot. Traffic=900Gbps

|  | Single layer Survivability | Multilayer Survivability | | |
| --- | --- | --- | --- | --- |
|  |  | double protection | LSP spare unprotected | interlayer BRS |
| Transit traffic | 107.5 Gbps | 52.5 Gbps | 52.5 Gbps | 52.5 Gbps |
| No. of lightpaths | 208 (105) | 226 (10) | 216 (10) | 216 (10) |
| No. of wavelengths | 596 | 505 | 490 | 480 (0) |
| Total cost | 5410 | 5399 (0.2%) | 5184 (4%) | 5154 (5%) |
| Optical layer cost | 1788 | 1515 (15%) | 1470 (18%) | 1440 (20%) |

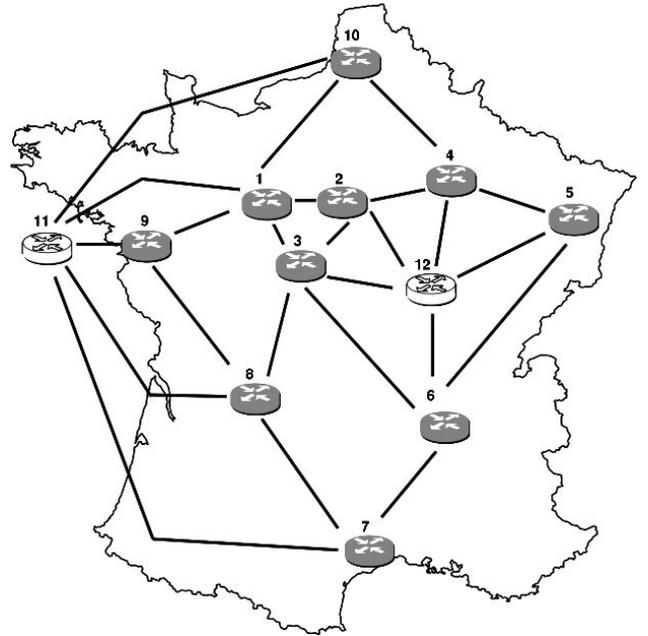

Fig. 6. The 12-node test network.

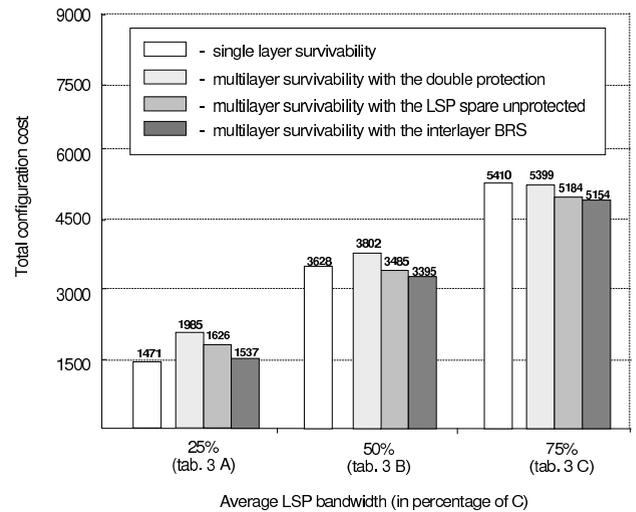

Fig. 7. Cost comparison of different survivability options.

The main difference between the single layer (SL) and the multilayer (ML) survivability is due to the fact that with the ML survivability only multihop LSPs are subject to protection routing and hence consume the MPLS spare capacity, whereas with the SL survivability both single- and multi-hop LSPs require spare resources of the MPLS layer. This results in a higher number of lightpaths carrying protection LSPs (see values in brackets in Table III(a)–(c)) and higher transit traffic obtained with the SL survivability. However, with the ML survivability protection lightpaths are added, so the total number of lightpaths is higher in this scenario. This in turn affects the total configuration cost as lightpaths are the most expensive network resources.

For low bandwidth LSPs, the SL survivability is the cheapest option, as depicted in Table III(a) and Fig. 7. Low bandwidth LSPs tend to be routed in multiple logical hops as they are groomed in intermediate routers to better fill the lightpath capacity. As a result, there exist relatively many multi-hop LSPs which are subject to protection routing with the ML survivability and more lightpaths are added (21) after protection LSP routing, despite a small total number of lightpaths. This makes the ML survivability the most expensive option for low bandwidth LSPs.

This relation changes for high-bandwidth LSPs, close to the lightpath capacity (Table III(c)). High bandwidth LSPs tend to be routed in single logical hops, i.e. on direct lightpaths, to minimize the transit traffic (note the small amount of transit traffic, as compared with the other traffic scenarios). Therefore, there exist relatively few multi-hop LSPs which are subject to protection routing with the ML survivability. As a result, much fewer lightpaths are added after the protection LSP routing (10), as compared with the SL survivability (105). This tendency and further savings in the wavelength use brought by the ML survivability (respectively 15%, 18% and 20% with different SCA methods) make this option the cheapest in this traffic scenario. The difference in the wavelength usage is due to the fact that with the SL survivability working and protection LSPs have disjoint routes in both network layers, whereas with the ML survivability working and protection lightpaths are disjoint only in the optical layer. As a result, lightpaths take longer routes with the SL survivability.

*2. Multilayer survivability - analysis of different SCA methods.*

Decreasing cost trends from "double protection" to "LSP spare unprotected" to "interlayer BRS" was expected as the spare resources of the MPLS layer are supported more and more efficiently by the optical layer. The lower total number of lightpaths achieved with the "LSP spare unprotected" method over the "double protection" is due to the fact that the lightpaths carrying protection LSPs (respectively 21, 20 and 10 with different traffic scenarios) are not protected. Savings in the wavelength usage stem from the fact that fewer lightpaths in total are routed in the physical topology (respectively 66 vs. 82, 148 vs. 160 and 216 vs. 226 for different traffic scenarios). Further savings are brought by the "interlayer BRS" method due to wavelength sharing among lightpaths carrying protection LSPs and protection lightpaths. One can see that most of the wavelengths used by the protection lightpaths are reused by the lightpaths carrying protection LSPs. Only, respectively, 4, 7, and 0 extra wavelengths are needed to accommodate protection LSPs within the optical layer for different traffic scenarios. This gives the reuse factor equal respectively to 84%, 80% and 100%.

*3. Impact of physical network characteristics and the cost structure.*

The difference between the SL and ML survivability is affected as well by the characteristics of the physical network topology (e.g. size, connectivity) and the assumed cost model. Fig. 8 shows the configuration cost of the SL survivability and the ML Interlayer BRS for different cost scenarios, as defined in the cost model (cf. Sec. IV-B, Table I). The total configuration cost, as well as the cost difference between both options are highly influenced by the cost of individual network components. The most expensive configurations result from using cost ratio CR_3, with the switching cost as a predominant factor. This is due to the fact that switching contributes both to the lightpath and wavelength cost. Note that cost ratio CR_3 accounts for opaque network architectures (OEO), where the cost of electronic switching matrices is still higher as compared with all-optical switches (OOO). Therefore, the configuration cost based on CR_3 is representative for OEO architectures; in particular, it demonstrates a higher CAPEX, as compared with all-optical networks (CR_1, CR_2).

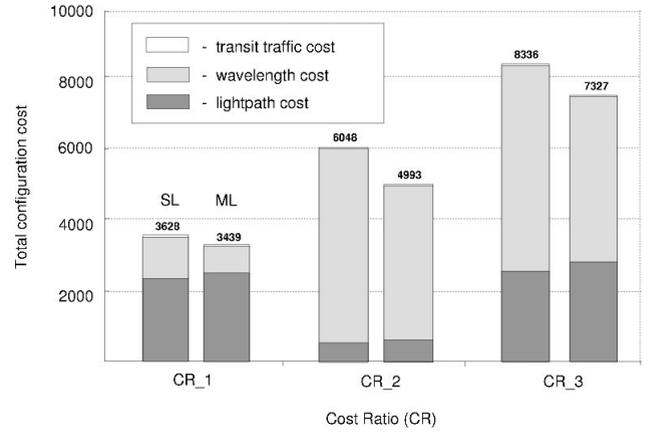

Fig. 8. SL vs. Interlayer BRS (ML) survivability for different cost scenarios. Traffic distribution and physical topology as of Fig. 6; average LSP bandwidth=0,5$C$; cost ratios (CR), as defined in Table 1.

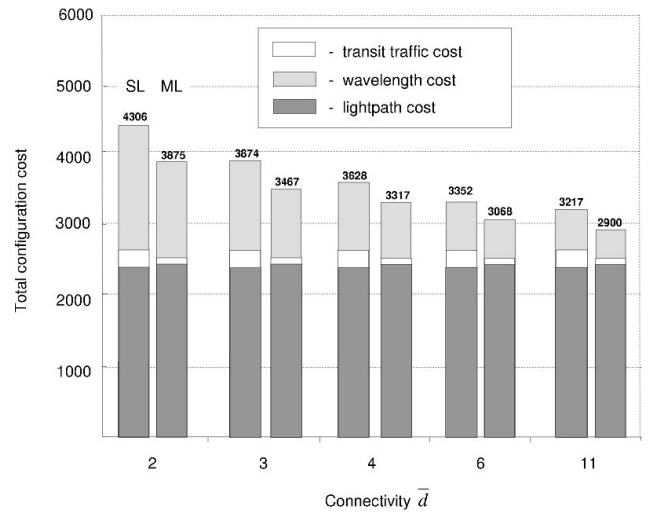

Fig. 9. SL vs. Interlayer BRS (ML) survivability for different physical topologies. Traffic matrix consisting of 126 LSPs; average LSP bandwidth=0,5$C$; 12-node physical network; cost ratio CR_1.

To verify the impact of the physical network topology on the network resources usage, the set of tests was carried out for randomly generated 12-node bi-connected network topologies with connectivity $\bar{d}$ varying from 2 (a ring) to 11 (a full mesh). For each network the same traffic matrix was used consisting of 126 LSPs, with the average LSP traffic equal to 0,5C. Results comparing the SL survivability and the ML Interlayer BRS, summarized in Fig. 9, bring us to the straightforward observation that sparse topologies consume more wavelengths to route the same number of lightpaths in the physical network. For a 12-node network the average lightpath length (measured in the physical hop count) varies from 3,33 ($\bar{d} = 2$) to 1 ($\bar{d} = 11$) for the lightpaths transporting working LSPs, and from 4,08 to 1,67 for the lightpaths carrying protection LSPs. The difference between the SL and ML survivability

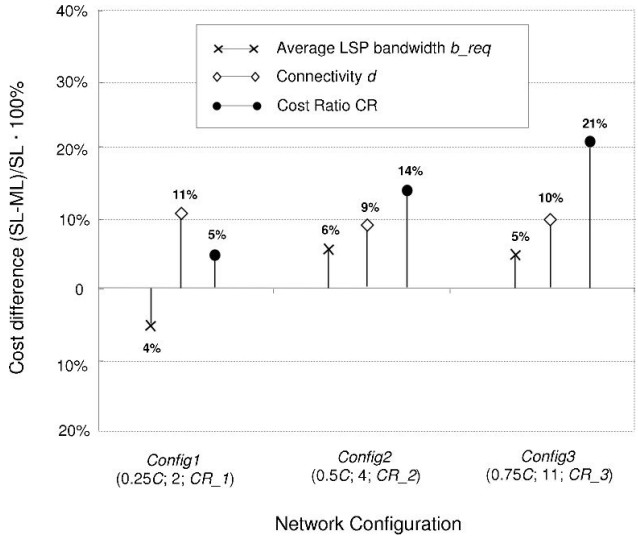

Fig. 10. Relative cost difference $\left(\frac{SL-ML}{ML} \cdot 100\%\right)$ between SL and Interlayer BRS (ML) survivability for different values of input parameters: LSP traffic granularity, network connectivity and cost.

remains fixed around 10% in favor of the ML option for all connectivity values.

Finally, we try to figure out which of the input parameters: (i) offered traffic characteristics (distribution, granularity), (ii) physical network topology (size, connectivity) or (iii) variations in the cost structure influence the most the survivability implementation. To verify the impact of these parameters on the survivability cost, a set of tests was carried out with changing one parameter within each test: average LSP bandwidth $\overline{b\_req}$ (0,25C, 0,5C, 0,75C), physical network connectivity $\overline{d}$ (2, 4, 11) and cost ratio CR (CR_1, CR_2, CR_3). Test results were then grouped into three configurations: $Config\_1$ ($\overline{b\_req} = 0,25C$; $\overline{d} = 2$; CR_1), $Config\_2$ ($\overline{b\_req} = 0,5C$; $\overline{d} = 4$; CR_2) and $Config\_3$ ($\overline{b\_req} = 0,75C$; $\overline{d} = 11$; CR_3), and for each configuration the cost difference between the SL and ML Interlayer BRS survivability was reported separately for each varying parameter. The results, summarized in Fig. 10, show that the relative cost difference varies the most for different cost ratios (from 5% to 21%) and different traffic granularities (from –4% to 6%), while it is almost constant for different physical network topologies. Based on that, we can say that the relative difference between various survivability options depends mainly on the offered traffic granularity and the assumed cost model, while the characteristics of the physical network have almost no impact on this. On the other hand, the physical network topology affects the amount of optical resources (i.e. wavelengths) used to route the lightpaths in the physical network, as reported in Fig. 9.

### 4. Sequential vs. integrated configuration method.

As a next step we tested the impact of the sequential vs. integrated configuration method on network resource usage and configuration cost. The results produced by both methods for different survivability options and the traffic matrix consisting of medium-size LSPs are summarized in Table IV and Fig. 11. In all cases we observed a gain in the wavelength use brought by the integrated method over the sequential one, while the MPLS layer resources (i.e. no. of established lightpaths and the transit traffic) were exactly the same. The difference is due to the fact that with the sequential approach the logical topology design and the lightpath routing are separated and the lightpaths are configured without the knowledge of the physical layer resources. As a consequence, the resulting logical topology is sub-optimal with respect to the wavelength usage and some lightpaths require longer physical routes. This effect is avoided with the integrated method where both processes are configured jointly. This allows the lightpaths to be optimally designed also in terms of the wavelength use (in fact, about 40% of lightpaths had different termination points when configured with the integrated method as compared to the sequential one). The difference is slightly higher for sparse physical network topologies (i.e. with low connectivity) which consume more wavelengths during the lightpath routing. For these networks logical topologies obtained with the sequential approach are more 'random' (i.e. less optimal) with respect to the lightpath routing which follows, and the resulting wavelength utilization. Similarly, the gain from using the integrated approach is higher for the configuration with working paths only (i.e. without survivability), as with the survivability implementation the wavelength routing is subject additionally to the protection routing constraints, which tighten the solution space and leave less room for optimization. To sum up, the most advantages from using the integrated approach are expected for sparse networks and for the working paths routing as opposed to the protection routing.

As the number of lightpaths remains the same for both approaches which constitute about 70% of the total configuration cost (cf. Fig. 9), the contribution of the integrated approach to the total cost savings is only 1.4%-3%. This rather small improvement brings us to the conclusion that the use of the sequential approach may be preferable, especially if taking into account much lower times spent optimization as compared with the integrated method (cf. Table 2).

## VI. SUMMARY AND FURTHER WORK

This study explored some design principles of MPLS over OTN architectures employing wavelength switching and targeted to a nationwide coverage. We used the ILP optimization combined with a cost model to dimension the network with the minimal resource usage and configuration cost. The obtained results are based on a cost model with actual technology pricing and representative traffic matrices.

We presented two approaches to the MPLS over OTN design and investigated various options for the survivability implementation. The comparative analysis between the single- and multilayer survivability shows the impact of the LSP traffic granularity, physical network structure and the assumed cost model on network resources usage and configuration cost, with the cost structure and offered traffic characteristics having the highest impact on them. Logical topology and the amount of transit traffic in routers are affected by the structure of the offered traffic: traffic matrices of high granularity (i.e. low bandwidth per LSP) result in sparse logical topologies, where LSPs are routed in multiple logical hops resulting in high transit traffic. The relatively high number of multi-hop LSPs increases the cost of the ML survivability, as compared with the SL one. On the other hand, high bandwidth

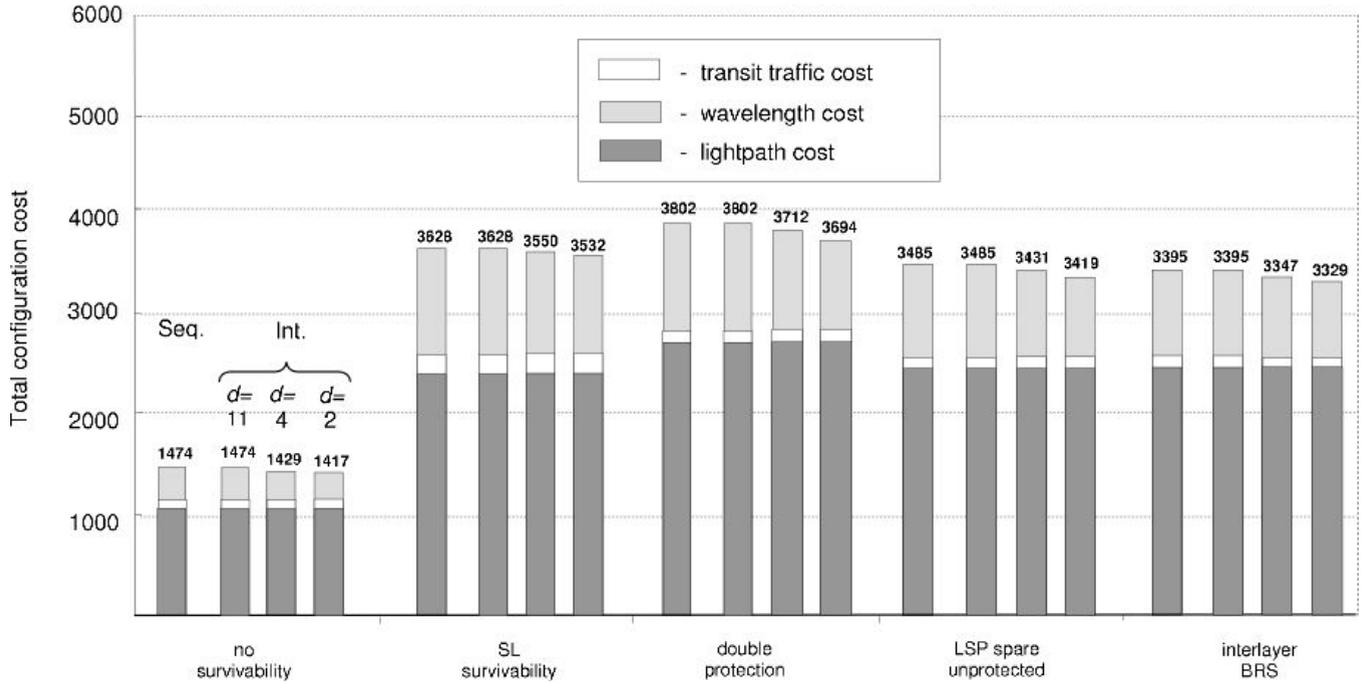

Fig. 11. Cost comparison of the sequential (Seq.) vs. integrated (Int.) configuration method for different values of the connectivity parameter $\overline{d}$.

TABLE IV
SEQUENTIAL VS. INTEGRATED CONFIGURATION METHODS FOR DIFFERENT SURVIVABILITY OPTIONS WITH THE AVERAGE LSP BANDWIDTH EQUAL TO 50% OF $C$, NETWORK CONNECTIVITY $\overline{d} = 4$ (CF. FIG. 6), AND COST RATIO $CR\_1$

| Survivability Option | Sequential | | | Integrated | | | Reduction | | |
|---|---|---|---|---|---|---|---|---|---|
| | Transit traffic (Gbps) | # of lightpaths | # of wavelengths | Transit traffic (Gbps) | # of lightpaths | # of wavelengths | Wavelengths | Optical layer cost | Total cost |
| No survivability | 77.5 | 64 | 108 | 77.5 | 64 | 93 | 15 | 14 % | 3 % |
| SL survivability | 262.5 | 143 | 329 | 262.5 | 143 | 303 | 26 | 8 % | 2 % |
| Double protection | 100 | 160 | 334 | 100 | 160 | 304 | 30 | 9 % | 2 % |
| LSP spare unprotected | 97.5 | 148 | 297 | 97.5 | 148 | 279 | 18 | 6 % | 1.5 % |
| Interlayer BRS | 97.5 | 148 | 267 | 97.5 | 148 | 251 | 16 | 6 % | 1.4 % |

LSPs, close to the lightpath capacity, result in dense logical topologies of many lightpaths with the LSPs routed in single logical hops and low transit traffic. The high number of lightpaths increases further the cost of the ML survivability, as compared with the SL one, as well as the overall cost of network configuration. On the other hand, physical network characteristics (size, connectivity) have the impact on the lightpath routing: sparse physical network topologies of low connectivity use more wavelengths for the lightpath routing and increase the configuration cost as compared with more dense networks. The total configuration cost is influenced a well by the cost of individual network components. The most expensive configurations result from using cost models with the switching cost as a predominant factor. Such cost structures are characteristic for OEO network architectures using electronic switching matrices and show the increased CAPEX of 'opaque' networks, as compared with all-optical ones.

We have found the integrated configuration method up to 9% more cost-efficient in terms of the wavelength use as compared with the sequential one, however, at the increase in the optimization problem complexity. On the other hand, as there is no effect of the configuration approach on the transit traffic and the number of lightpaths, the total cost reduction brought by the integrated approach is no more than 3%. This rather small cost improvement reveals the usefulness of the sequential approach based on decomposition to improve the optimization time and, consequently, to increase the size of problems to be handled, at the expense of a relatively small drop in the solution quality.

Authors believe that the proposed ILP-based framework could be used as well to address dynamic network conditions,

where LSP demands are not all known in advance. It would be interesting to use the ILP method in the context of dynamic reconfiguration to determine efficient re-optimization (traffic engineering) policies, i.e. which involve few changes to adapt to new network state, (possibly) without deteriorating the current optimum. Nevertheless, the use of heuristics is inevitable to attack larger networks. Some well-known heuristics for computing a set of disjoint paths, such as, for example, *branch exchange* described in [9] or others, based on Evolutionary Algorithms and presented in [8, 20-21] can be used within the presented design framework to improve the scalability of the proposed solutions.

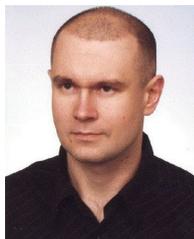

**Wojtek Bigos** received the M.S. degree ('01) from the AGH University of Technology in Krakow, Poland and the Ph.D. degree ('06) from the University of Rennes 1, France, in co-operation with the R&D Division of France Télécom. Currently he is a network engineer at Nokia Siemens Networks Development Centre in Wroclaw, Poland. His professional interests include network planning and optimization incl. dimensioning algorithms, traffic flow control methods and performance optimization of access and core network architectures.

**Bernard Cousin** is a Professor of Computer Science at the University of Rennes 1, France. Prior to joining the University of Rennes, he was assistant professor at Bordeaux. He is an elected member of the board of the computer science department. He is president of the scientific committee on computer science for the university of Rennes 1. He is, currently, a member of IRISA (a CNRS-INRIA-University-INSA joint research laboratory in computing science located at Rennes). More specifically, he is at the head of a research group on networking. He is the co-author of a network technology book: *IPV6* (Fourth edition, O'Reilly, 2006) and has co-authored a few IETF drafts in the areas of Explicit Multicasting and Secure DNS. His research interests include high speed networks, traffic engineering, multicast routing, network QoS management, network security, dependable networking, and distributed and multimedia applications. Bernard Cousin received his Ph.D. degree in computer science from the University of Paris 6, in 1987.

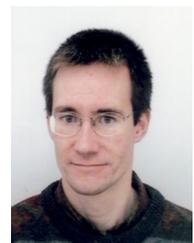

**Stéphane Gosselin** was born in 1968, received a general science degree from Ecole Polytechnique, a telecommunications degree from École Nationale Supérieure des Télécommunications, and a research master in Optoelectronics. He joined France Telecom CNET at Lannion in 1993, where he first studied fast optically-addressed spatial light modulators and fast free-space optical switching systems. From 1997, he has been working on WDM networks, optical cross-connects and optical transmission. He led an R&D group of France Telecom R&D on optical transport network technologies from 1999 to 2003, and has been leading France Telecom long term research on optical networks since 2003. He was/is involved in several French and European (ACTS/DEMON, IST/TOPRATE, IST/NoE/e-Photon/One) projects dealing with optics. He authored four international patents and about 40 national or international papers or communications. He has been a member of technical program committee of ECOC since 2004.

**Morgane Le Foll's** biography was not available at the time of publication.

**Hisao Nakajima** is doctor in engineering. Since 1986, he has been with France Télécom R&D and has worked as a senior research engineer in optoelectronics and in optical networking. He participated in the RACE 1027 (coherent optical communication system), RACEII/ATMOS (ATM optical switching) projects, the EURESCOM P-1012 FASHION (ASON) and P-1116 SCORPION (IP/ASON) projects, the IST/NOBEL project (NGN) and the French RNRT/Rythme project (hybrid optical networks). He is currently involved in the IST/NOBEL phase 2 project and in the French RNRT/ECOFRAME project. He authored or co-authored more than 90 technical papers and 14 patents. His current interests include optical control plane and interworking issues in IP/optical networks and optical burst switched networks.